\documentclass{PoS}
\usepackage{cite}
\usepackage{amsmath}  
\usepackage{amssymb} 
\usepackage{dsfont}
 \usepackage{multirow}
\usepackage{psfrag,scalefnt}
\usepackage{paralist}

\usepackage{color}
\let\normalcolor\relax

\newcommand{\vub}{|V_{ub}|}

\newcommand{\tev}{\, {\rm TeV}}

\newcommand{\be}{\begin{equation}}
\newcommand{\ee}{\end{equation}}
\newcommand{\bea}{\begin{eqnarray}}
\newcommand{\eea}{\end{eqnarray}}

\newcommand{\ba}{\begin{array}}
\newcommand{\ea}{\end{array}}

\newcommand{\ord}{{\cal O}}

\def\kpn{K^+\rightarrow\pi^+\nu\bar\nu}

\def\klpn{K_{L}\rightarrow\pi^0\nu\bar\nu}

\title{Towards the Identification of New Physics through Correlations between Flavour Observables}

\ShortTitle{Towards the Identification of New Physics}

\author{\speaker{Andrzej J. Buras} \addtocounter{footnote}{1} \\
TUM-IAS, Lichtenbergstr. 2a, D-85748 Garching, Germany \\
Technical University Munich, Physics Department, D-85748 Garching, Germany,\\
E-mail: \email{Andrzej.Buras@ph.tum.de}}

\abstract{We emphasize the power of correlations between flavour observables in the search for New Physics and  identify 
a number of correlations that could allow to discover New Physics even if it would appear at the level of $20\%$ of the Standard Model contributions.
After presenting the simplest correlations in CMFV and $U(2)^3$ models 
we address the recent data on $B_{s,d}\to\mu^+\mu^-$ and the 
anomalies in $B_d\to K^*\mu^+\mu^-$ in the context of $Z^\prime$-models and 
SM $Z$, both with flavour violating neutral couplings.
A strategy consisting of twelve steps which concentrate on theoretically clean observables, to be measured in this decade, could one day allow us to reach the Zeptouniverse. The related 
DNA charts based on the correlations between enhancements and suppressions of various observables in a given New Physics scenario relative to the Standard Model predictions allow a transparent distinction between various extensions of this model. The present text is the extended version of the talk to be published in 
the proceedings of this symposium.
}

\FullConference{The 2013 EPS Symposium on High Energy Physics\\
                 July 17-24, 2013\\
                 Stockholm, Sweden}

\begin{document}

\section{Overture}
In spite of tremendous efforts of experimentalists and theorists to find 
New Physics (NP) beyond the Standard Model (SM), no clear indications for NP beyond dark matter, neutrino masses and matter-antimatter asymmetry in the universe have been observed. Yet, the recent discovery of a Higgs-like particle and the overall agreement of the SM with the present data shows that our general approach of describing physics at very short distance scales with the help of exact (QED and QCD) and spontaneously broken (for weak interactions) gauge theories is correct.

As the SM on the theoretical side is not fully satisfactory and the three 
NP signals mentioned above are already present, we know that some new particles 
and new forces have to exist, hopefully within energy scales being presently directly explored by the LHC or not far above them.
The upgrade in the energy of the LHC, the upgrade of the LHCb, SuperKEKB and dedicated 
Kaon physics  experiments at CERN, J-PARC and Fermilab, as well as improved 
measurements of lepton flavour violation (LFV), electric dipole moments (EDMs) and 
$(g-2)_{\mu,e}$ will definitely shed light on the question whether NP is present 
below, say, $100\tev$. However in the coming decades to go beyond $10\tev$  will require the study of very rare
processes. These are
 in particular  flavour violating and CP-violating rare 
decays of mesons, EDMs, LFV and $(g-2)_{\mu,e}$. As this is an indirect 
search for NP one has to develop special strategies to reach  the Zeptouniverse, that is scales as short as $10^{-21}{\rm m}$ or equivalently energy scales as high as several hundreds of TeV.
The present talk discusses some of  such strategies developed in my group at the Technical University in Munich during last ten years. They are summarized in 
\cite{Buras:2013ooa}\footnote{The updated version of this review will appear in  October this year.}. 

\section{Main Strategy}
The identifcation of  NP through rare processes, that is through quantum fluctuations, will require
\begin{itemize}
\item
many precise measurements of many observables and precise theory,
\item
intensive studies of correlations between many observables in a given extension of the SM with the goal to identify patterns of deviations 
from the SM expectations characteristic for this extension,
\item
intensive studies of  correlations between low energy precision measurements and the measurements at the highest available energy, that is in the coming decades the measurements in proton-proton collisions at the LHC.
\end{itemize}

Now in the search for NP 
 one distinguishes between {\it bottom-up} 
and {\it top-down} approaches. In my view both approaches should be persued 
but I think that 
in the context of flavour physics and simultaneous 
 exploration of 
short distance physics both through LHC and high precision experiments the 
top-down approach is more powerful. Here are my arguments.

In the  bottom-up approach
one constructs effective field theories involving 
only light degrees 
of freedom including the top quark and Higgs boson in which the structure of the effective 
Lagrangians is governed by the symmetries of the SM and often other 
hypothetical symmetries. This approach is rather powerful in the case of
electroweak precision 
studies and definitely teaches us something about $\Delta F=2$ 
transitions. In particular lower bounds on NP scales depending on the 
Lorentz structure of operators involved can be derived from the data 
\cite{Bona:2007vi,Isidori:2010kg}.
However, except for the case of  minimal flavour violation (MFV) and closely related 
approaches based on flavour symmetries, the bottom-up approach ceases, 
in my view, to be useful in $\Delta F=1$ decays, 
because of very many operators that are allowed to appear
in the effective Lagrangians with coefficients that are basically 
unknown \cite{Buchmuller:1985jz,Grzadkowski:2010es}. In this 
approach then the correlations between various $\Delta F=2$ and $\Delta F=1$ 
observables in $K$, $D$, $B_d$ and $B_s$ systems are either not visible or 
very weak, again except MFV and closely related approaches. Moreover 
the correlations between flavour violation in low energy processes and 
flavour violation in high energy processes 
are lost. Again MFV is among few exceptions.

On the other hand
in the top-down 
approach one constructs first 
a specific model with heavy degrees of freedom. For high energy processes,
where the energy scales are of the order of the masses of heavy particles 
one can directly use this ``full theory'' to calculate various processes 
in terms of the fundamental parameters of a given theory. For low energy 
processes one again constructs the low energy theory by integrating out 
heavy particles. The advantage over the bottom-up approach is that now the 
Wilson 
coefficients of the resulting local operators are calculable in terms of 
the fundamental parameters of this theory. In this manner correlations between 
various observables belonging to different mesonic systems and correlations 
between low energy and high-energy observables are possible. Such correlations 
are less sensitive to free parameters than individual observables and 
represent patterns of flavour violation characteristic for a given theory. 
These correlations can in some models differ strikingly from the ones of 
the SM and of the MFV approach.

\begin{figure}[!tb]
\centering
\includegraphics[width=0.65\textwidth]{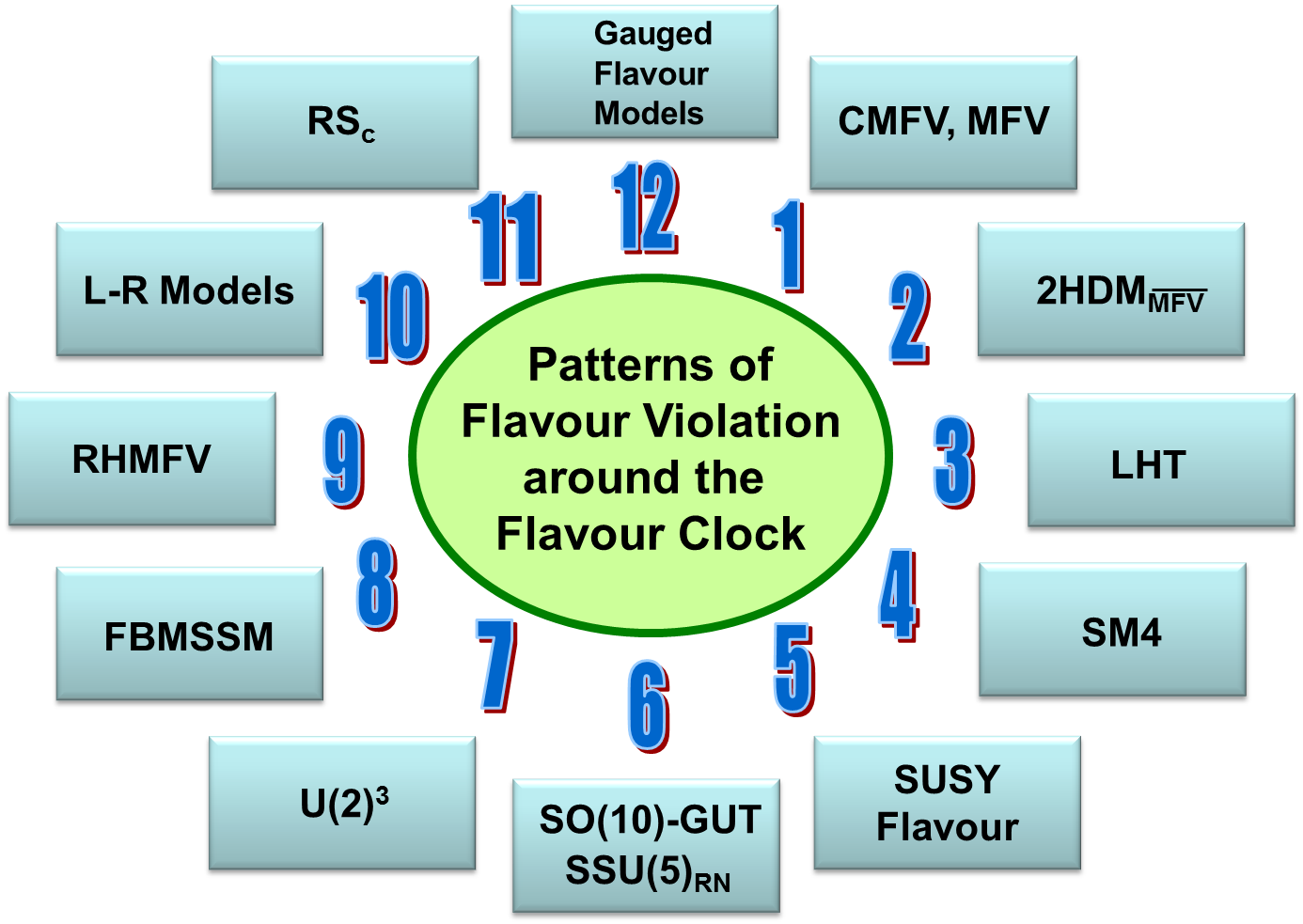}
\caption{\it Studing Multitude of Extensions of the Standard Model.}\label{Fig:2}~\\[-2mm]\hrule
\end{figure}

Having the latter strategy in mind I have in the last ten years
investigated together with my 
young collaborators 
flavour violating and CP-violating processes 
in a multitude of models. The names of models analyzed by us until June 2012 are collected in Fig.~\ref{Fig:2}. A summary of these studies with brief descriptions of all these models can be found in  \cite{Buras:2010wr,Buras:2012ts}. Here, I will concentrate on most recent analyses that have been performed after the second of these two reviews and are not shown in Fig.~\ref{Fig:2}.  They are 
reviewed in \cite{Buras:2013ooa}.


\begin{figure}[!tb]
 \centering
\includegraphics[width = 0.6\textwidth]{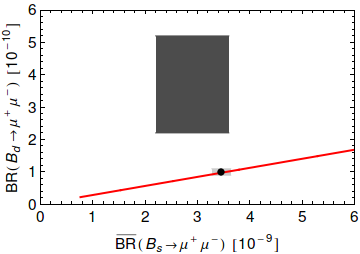}
\caption{\it $\mathcal{B}(B_d\to\mu^+\mu^-)$ vs $\overline{\mathcal{B}}(B_s\to\mu^+\mu^-)$ 
  in models 
with CMFV. SM is represented by the light grey area with black dot. Dark gray
 region: Overlap of  exp 1$\sigma$
 ranges for
 $\overline{\mathcal{B}}(B_s\to\mu^+\mu^-) = (2.9\pm0.7)\cdot 10^{-9}$ and 
$\mathcal{B}(B_{d}\to\mu^+\mu^-) =(3.6^{+1.6}_{-1.4})\times 10^{-10}$. From 
\cite{Buras:2013ooa}.}\label{fig:BdvsBs}~\\[-2mm]\hrule
\end{figure}

\section{Simplest Correlations}
I would like first to recall few correlations that are very simple and 
presently rather relevant. 
The first two are the ones in models with constrained Minimal Flavout Violation 
(CMFV)  \cite{Buras:2000dm,Buras:2003jf}
\begin{equation}\label{CMFV5}
 \frac{\mathcal{B}(B_s\to\mu^+\mu^-)}{\mathcal{B}(B_d\to\mu^+\mu^-)}=
 \frac{\tau({B_s})}{\tau({B_d})}\frac{m_{B_s}}{m_{B_d}}
 \frac{F^2_{B_s}}{F^2_{B_d}}
 \left|\frac{V_{ts}}{V_{td}}\right|^2,
 \end{equation}
and \cite{Buras:2003td}\footnote{As emphasized in \cite{Hurth:2008jc} the 
dependence of the ratio of branching ratios in (\ref{CMFV5}) on the elements 
of the CKM is more general than CMFV and applies to MFV at large \cite{Chivukula:1987py,Hall:1990ac,D'Ambrosio:2002ex}.}
\be\label{CMFV6}
 \frac{\mathcal{B}(B_{s}\to\mu^+\mu^-)}{\mathcal{B}(B_{d}\to\mu^+\mu^-)}
 =r\frac{\hat B_{d}}{\hat B_{s}}
 \frac{\tau( B_{s})}{\tau( B_{d})} 
 \frac{\Delta M_{s}}{\Delta M_{d}}=r\,(34.3\pm0.8),  \qquad 
\frac{\hat B_{d}}{\hat B_{s}}=0.99\pm0.02
\ee
where the departure of $r$ from unity measures effects which go beyond CMFV. 
This {\it golden} relation  between $\Delta M_{s,d}$ and $B_{s,d}\to\mu^+\mu^-$ 
does not 
 involve $F_{B_q}$ and CKM parameters and consequently contains 
 smaller hadronic and parametric uncertainties than (\ref{CMFV5}). It involves
 only measurable quantities except for the ratio $\hat B_{s}/\hat B_{d}$
  that is known from  lattice calculations with impressive 
accuracy of roughly $\pm 2\%$ \cite{Carrasco:2013zta} as given in (\ref{CMFV6})\footnote{This result is not included in the recent FLAG
update which quotes
$0.95\pm0.10$.}. 
 Consequently the r.h.s of this equation is already  rather precisely 
 known and this precision should be improved within this decade. 
 This would allow to identify possible NP in 
 $B_{s,d}\to \mu^+\mu^-$ decays and also in $\Delta M_{s,d}$  even if it was only at the level of $20\%$ of 
 the SM contributions. This is rather unique in the quark flavour physics and 
only the decays $\kpn$ and $\klpn$ can compete with this precision.

Indeed, the most recent data  on these very rare 
 decays  from LHCb and CMS collaborations give first indications that NP 
 contributions to $B_s\to\mu^+\mu^-$  are much smaller than the 
SM contribution
 and in particular the relation (\ref{CMFV6}) but also (\ref{CMFV5}) could 
 turn out to be an important tool in the coming years to identify NP. 
On the other hand the data on $B_d\to\mu^+\mu^-$ exhibit some departure 
from SM expectations but we have to wait for improved data in order to see 
whether NP is here at work. 
We compare the relation (\ref{CMFV6}) with present data in  Fig.~\ref{fig:BdvsBs}, where we included $\Delta\Gamma_s$ effects in $B_s\to\mu^+\mu^-$ as discussed below. We will soon investigate what kind of NP could give a better description of the data than it is presently the case of CMFV.

While experimentalists from CMS and LHCb should be congratulated on the 
measurements on such low branching ratios, their result for $B_s\to\mu^+\mu^-$ 
has been predicted by theorists more than a decade ago.
The first NLO-QCD calculation of these decays has been performed 20 years 
ago in \cite{Buchalla:1993bv}. In contrast to what is stated usually in the literature,  the most important result of this paper 
was not the reduction of the scale uncertainty due to the choice of the scale 
in $m_t$ but the inclusion of a factor of two in the branching ratios for $B_{s,d}\to\mu^+\mu^-$ which was missed in the previous literature.  This factor of two can  be 
appreciated for the first time this year. Indeed, the most recent predictions in the SM  \cite{Buras:2012ru,Buras:2013uqa}  and the 
most recent averages from LHCb \cite{Aaij:2013aka} and CMS 
\cite{Chatrchyan:2013bka} are given as follows:
\be\label{LHCb2}
\overline{\mathcal{B}}(B_{s}\to\mu^+\mu^-)_{\rm SM}= (3.56\pm0.18)\cdot 10^{-9},\quad
\overline{\mathcal{B}}(B_{s}\to\mu^+\mu^-) = (2.9\pm0.7) \times 10^{-9}, 
\ee
\be\label{LHCb3}
\mathcal{B}(B_{d}\to\mu^+\mu^-)_{\rm SM}=(1.05\pm0.07)\times 10^{-10}, \quad
\mathcal{B}(B_{d}\to\mu^+\mu^-) =(3.6^{+1.6}_{-1.4})\times 10^{-10}. \quad
\ee
The ``bar'' in the case of $B_{s}\to\mu^+\mu^-$ indicates that $\Delta\Gamma_s$ 
effects \cite{DescotesGenon:2011pb,DeBruyn:2012wj,DeBruyn:2012wk} have been taken into account. These two branching ratios are related through
\cite{DeBruyn:2012wk}
\be
\label{Fleischer1}
\mathcal{B}(B_{s}\to\mu^+\mu^-) =
r(y_s)~\overline{\mathcal{B}}(B_{s}\to\mu^+\mu^-), 
\ee
where 
\be\label{rys}
r(y_s)\equiv\frac{1-y_s^2}{1+\mathcal{A}^{\mu^+\mu^-}_{\Delta\Gamma} y_s}, \qquad
	y_s\equiv\tau_{B_s}\frac{\Delta\Gamma_s}{2}. 
\end{equation}
The observable $\mathcal{A}^{\mu^+\mu^-}_{\Delta\Gamma}$ can be extracted from the untagged time-dependent studies and generally depends on NP but knowing it experimentally allows to determine $\mathcal{B}(B_{s}\to\mu^+\mu^-)$ which is usually 
calculated by theorists. In the SM and CMFV we have $\mathcal{A}^{\mu^+\mu^-}_{\Delta\Gamma}=1$ and the inclusion of $\Delta\Gamma_s$ effects  rescales the branching ratio $\mathcal{B}(B_{s}\to\mu^+\mu^-)$ upwards. The amount of this rescaling depends on the experimental value of $y_s$. While at the time of the 
analyses in \cite{Buras:2012ru,Buras:2013uqa} one had $y_s=0.088\pm0.014$, 
the most recent value is $y_s=0.062\pm0.009$ \cite{Amhis:2012bh}. This  changes the SM value 
in (\ref{LHCb2}) to $3.46$ but we will not do it here as such small modifications should be included together with complete NLO electroweak corrections (Bobeth, Gorbahn and Stamou) and NNLO QCD corrections (Hermann, Misiak and Steinhauser)
 that should appear in the arxiv soon.

Clearly in the case of $B_d\to\mu^+\mu^-$ 
large deviations from SM prediction are still possible. But in the case of
$\overline{\mathcal{B}}(B_{s}\to\mu^+\mu^-)$
deviations by more than $30\%$ from its SM value seem rather unlikely. Yet, 
the reduction of the error in the SM prediction down to $3-4\%$ is still possible and this would allow to see NP at the level of $20\%$ provided the 
measurements improve.

We observe that while the data for $\overline{\mathcal{B}}(B_{s}\to\mu^+\mu^-)$ 
are by $1\sigma$  lower than the SM prediction, the data on $\mathcal{B}(B_{d}\to\mu^+\mu^-)$ are by $1.9\sigma$ above its SM value. Removing the $\Delta\Gamma_s$ effect by means of (\ref{Fleischer1}) from the experimental value for $\overline{\mathcal{B}}(B_{s}\to\mu^+\mu^-)$  we find 
\be\label{rexp}
r_{\rm exp}=0.22\pm 0.11
\ee
to be compared with $r=1$ in CMFV.
Even if  in view of large experimental uncertainties one cannot claim that  NP is at work here, the plot  in Fig.~\ref{fig:BdvsBs} invites us to investigate whether the simplest 
models could cope with the future more precise experimental results in which the central values of the branching ratios  in (\ref{LHCb2}) and (\ref{LHCb3})
would not change by much.

In CMFV and MFV at large \cite{D'Ambrosio:2002ex}, that are both 
based on the $U(3)^3$ flavour symmetry, the measurement of   
the mixing induced asymmetry  $S_{\psi K_S}$ together with the unitarity 
of the CKM implies that the analogous asymmetry in the $B_s^0-\bar B_s^0$ 
system, $S_{\psi\phi}$, is very small: $0.036\pm0.002$. Presently the data 
give
\be\label{SDATA}
S_{\psi K_S}= 0.679\pm 0.020,\qquad S_{\psi\phi}= -(0.04^{+0.10}_{-0.13})
\ee
and although $S_{\psi\phi}$ is found to be small   \cite{Amhis:2012bh} it can still significantly 
differ from its SM value, in particular if it had negative sign\footnote{Our definition 
of $S_{\psi\phi}$ differs by sign from the one used by LHCb and HQAG.}.

\begin{figure}[!tb]
 \centering
\includegraphics[width = 0.6\textwidth]{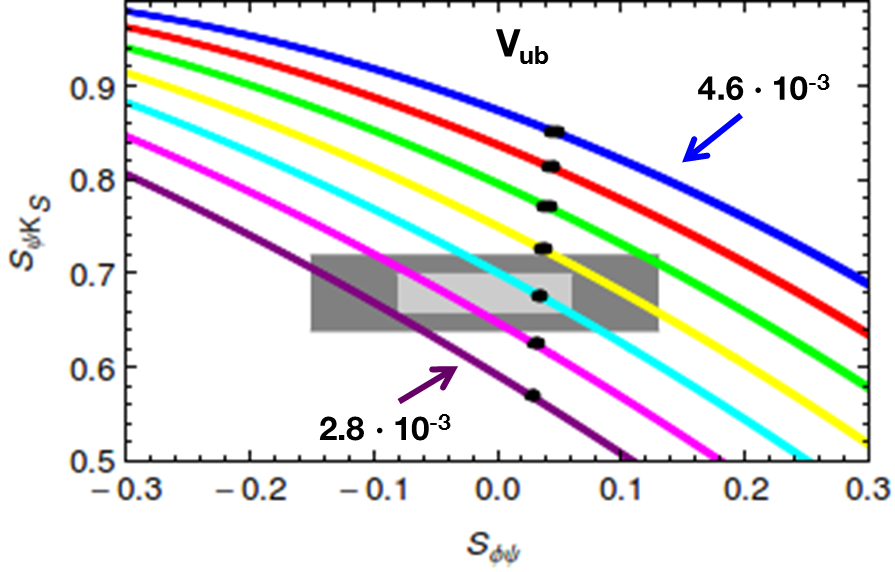}
\caption{ \it $S_{\psi K_S}$ vs. $S_{\psi \phi}$ in  models with 
$U(2)^3$ symmetry for different values of $\vub$ and $\gamma\in[58^\circ,78^\circ]$. From top to bottom: $\vub =$ $0.0046$ (blue), $0.0043$
(red), $0.0040$ (green),
$0.0037$ (yellow), $0.0034$ (cyan), $0.0031$ (magenta), $0.0028$ (purple). Light/dark gray: experimental $1\sigma/2\sigma$ region.
}\label{fig:SvsS}~\\[-2mm]\hrule
\end{figure}

If this indeed turned out to be the case, one possible solution would be 
to decrease the flavour symmetry down to $U(2)^3$, the NP scenario studied 
in particular in \cite{Barbieri:2011fc,Barbieri:2012uh,Crivellin:2011fb}.
 As  pointed out in \cite{Buras:2012sd} in the simplest versions of these models in which this symmetry 
is broken minimally, there is a stringent triple correlation
 $S_{\psi K_S}-S_{\psi\phi}-|V_{ub}|$  that  constitutes 
an important test of this NP scenario. 
 We 
show this correlation in Fig.~\ref{fig:SvsS}  for  $\gamma$ between $58^\circ$ and $78^\circ$. The latter dependence is very weak and is represented by the thickness of the lines.  Note that in a $U(2)^3$ symmetric world, $\vub$ could be 
determined with very small hadronic uncertainties by simply measuring $S_{\psi\phi}$ and $S_{\psi K_S}$. However, it is more interesting to extract $\vub$ from tree level decays and check whether this triple correlation is satisfied. 

But now comes an important point. In 
this simple scenario the relation (\ref{CMFV6}) is still valid \cite{Buras:2012sd} even if the branching ratios and $\Delta M_{s,d}$ can all differ  from their SM values. This means that if the experimental grey area in Fig.~\ref{fig:BdvsBs} will not move 
 and it will decrease in the future our world is either not $U(2)^3$ symmetric 
or the breakdown of this symmetry is more involved. We 
will then have to look for other alternatives. One of the simplest alternatives 
that can cope with this challenge are models with tree level FCNCs which we will discuss next.

\section{Correlations between Flavour Observables in Models with Tree Level FCNCs}
During the last year we have studied flavour observables in models in which 
FCNC processes are mediated at tree-level by neutral gauge bosons \cite{Buras:2012dp,Buras:2012jb,Buras:2013td} and neutral scalars or pseudoscalars \cite{Buras:2013uqa,Buras:2013rqa}. While such processes have been studied in the literature for the last three decades\footnote{A review of $Z^\prime$ models can be found in  \cite{Langacker:2008yv} and 
other recent studies in these models have been presented in 
\cite{Barger:2009qs,Fox:2011qd,Altmannshofer:2011gn,Altmannshofer:2012az,Dighe:2012df,Sun:2013cza,Descotes-Genon:2013wba,Altmannshofer:2013foa,Gauld:2013qba}.}, we still could contribute to this field by 
identifying certain correlations between several flavour observables that have 
not been presented in the past. Moreover, in \cite{Buras:2012fs} we have calculated for the first time the complete  
NLO-QCD corrections to tree-level contributions of colourless gauge bosons and 
scalars to $\Delta F=2$ transitions. The corresponding calculations for non-leptonic $\Delta F=1$ transitions have been presented in \cite{Buras:2012gm}.

The structure of such NP contributions is very simple.
A tree level contribution to a $\Delta F=2$ transition, like particle-antiparticle mixing, mediated by a gauge boson $Z^\prime$ is described by the amplitude
\be\label{FCNC1}
\mathcal{A}(\Delta F=2)=a \bar\Delta_B^{ij}(Z^\prime)\bar\Delta_C^{ij}(Z^\prime), 
\qquad \bar\Delta_B^{ij}(Z^\prime)=\frac{\Delta_B^{ij}(Z^\prime)}{M_{Z^\prime}},
\ee
where $\Delta_{B,C}^{ij}$  with $(B,C)=(L,R)$ are left-handed or right-handed 
couplings of $Z^\prime$ to quarks with $(i,j)$ equal 
to $(s,d)$, $(b,d)$ and $(b,s)$ for $K^0$, $B^0_d$ and $B^0_s$ meson system, 
respectively. The overall flavour independent factor $a$ is a numerical constant that generally depends 
on $L$ and $R$ but we suppress this dependence.
If we assume that only left-handed or right-handed couplings 
are present or that left-handed and right-handed couplings are either equal 
to each other or differ by sign, then this amplitude for a fixed 
$(i,j)$ is described only by two parameters, the magnitude and the phase of 
the reduced coupling $\bar\Delta_{B}^{ij}$.

On the other hand a tree-level amplitude for a $\Delta F=1$ transition like 
a leptonic or semi-leptonic decay of a meson with $\mu\bar\mu$ in the final 
state has the structure
\be\label{FCNC2}
\mathcal{A}(\Delta F=1)=b\bar\Delta_B^{ij}(Z^\prime)\bar\Delta_D^{\mu\bar\mu}(Z^\prime), \qquad 
\bar\Delta_D^{\mu\bar\mu}(Z^\prime)=
\frac{\Delta_D^{\mu\bar\mu}(Z^\prime)}{M_{Z^\prime}},
\ee
with $\bar\Delta_B^{ij}(Z^\prime)$ being the same quark couplings as in (\ref{FCNC1}) and $b$ is again an overall factor. $D=(A,V)$ distinguishes between 
 axial-vector 
and vector coupling to muons.
Clearly the same formulae with different values of couplings and the factors 
$a$ and $b$ apply to a tree-level 
exchange of $Z$, a heavy pseudoscalar $A$, a heavy scalar $H$ and the Higgs boson. In the latter case the contributions to rare decays are very small once 
the flavour violating couplings are constrained through mixing because of 
the small leptonic couplings with a possible exception of the $\tau$.
In what follows we do not assume that this suppression is valid for 
heavy pseudoscalar and scalar and that their couplings to leptons are not proportional to lepton masses as is the case of the SM Higgs. Moreover it should 
be emphasized that in a particular NP scenario the FCNC couplings in question 
could be generated first  at the one-loop level. Also in this case 
the formulae above would apply but then one should check whether genuine 
loop contributions to FCNC processes in this model are equally important or 
even more important. In what follows we assume that such contributions are 
subleading.

Now we can constrain the $\Delta_B^{bs}(Z^\prime)$ couplings by the data on 
$\Delta M_s$ and 
the CP-asymmetry $S_{\psi\phi}$ and the couplings $\Delta_B^{bd}(Z^\prime)$ by the  data on 
$\Delta M_d$ and the CP-asymmetry $S_{\psi K_S}$. In the case of $\Delta_B^{sd}(Z^\prime)$ we have mainly $\varepsilon_K$ to our disposal as $\Delta M_K$ 
being subject to significant hadronic uncertainties provides much weaker 
constraint than $\varepsilon_K$   in the models in question.

Once these constraints on the magnitude and the phase of 
new couplings are imposed and the allowed values 
are used for the predictions for rare decays it is evident that correlations 
between various observables are present. It is particularly interesting that
the pattern of these correlations depends on whether a gauge boson,  a scalar 
or pseudoscalar mediates the FCNC transition. As the scalar contributions cannot interfere with SM contributions, only enhancements of branching ratios are possible in this case. A tree-level gauge boson contribution and pseudoscalar 
contribution interfer generally with the SM contribution but the resulting 
correlations between observables have different pattern because of the $i$ 
in the coupling 
$i\gamma_5$ of a pseudoscalar to leptons. We refer for detailed analytic explanation of these differences to \cite{Buras:2013rqa}.

\begin{figure}[!tb]
\centering
 \includegraphics[width= 0.45\textwidth]{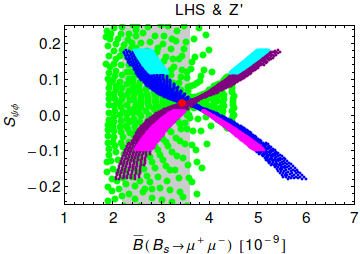}
 \includegraphics[width= 0.45\textwidth]{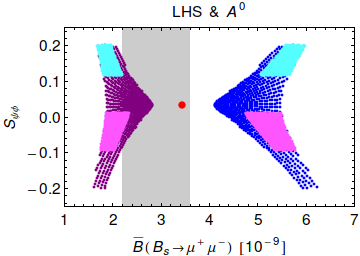}
\includegraphics[width= 0.45\textwidth]{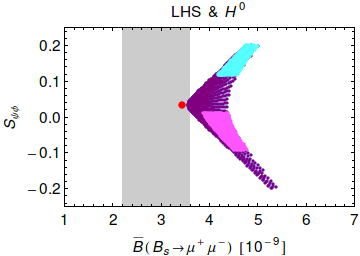}
\caption{\it  $S_{\psi\phi}$ versus $\overline{\mathcal{B}}(B_s\to\mu^+\mu^-)$ 
in different tree-level NP scenarios as explained in the text with
$M_{Z^\prime} =M_{A^0}= M_{H^0}=1~$TeV.
Gray region:
exp 1$\sigma$ range
$\overline{\mathcal{B}}(B_s\to\mu^+\mu^-) = (2.9\pm 0.7)\cdot 10^{-9}$.  Red point: SM central
value.}\label{fig:BsmuvsSphiZprimeA}~\\[-2mm]\hrule
\end{figure}

 In Fig.~\ref{fig:BsmuvsSphiZprimeA} we show the correlation between $S_{\psi\phi}$ and $\overline{\mathcal{B}}(B_{s}\to\mu^+\mu^-)$ in the case of pure left-handed $Z^\prime$, pseudoscalar 
and scalar couplings. The blue and magenta regions correspond to two solutions 
for the phase of the involved coupling $\Delta^{bs}_L$, differing by $180^\circ$,  which cannot be distinguished 
by $\Delta M_s$ and $S_{\psi\phi}$ alone. In the $Z^\prime$ case they correspond 
to correlation and anticorrelation between $S_{\psi\phi}$ and $\overline{\mathcal{B}}(B_{s}\to\mu^+\mu^-)$.  The smaller 
cyan and purple regions are obtained when $U(2)^3$ symmetry is imposed on the 
couplings. The same colour coding is used in the $A^0$ and $H^0$ case. The 
gray region corresponds to one $\sigma$ in (\ref{LHCb2}) and (\ref{SDATA})
and the green region 
is allowed by other data on $b\to sl^+l^-$ transitions like $B\to K^*ll$. Further distinction between different regions can be obtained by studying other observables like CP-asymmetry $S_{\mu\mu}^s$ in $B_{s}\to\mu^+\mu^-$ and the transitions $b\to s\nu\bar\nu$ but the correlation with the latter decays requires the 
relative size of muon and neutrino couplings. There is no space to present these correlations here. They can be found in \cite{Buras:2013rqa}. In particular simultaneous study of $B\to K (K^*)\nu\bar\nu$ and $B\to X_s\nu\bar\nu$ can provide information about the importance of right-handed couplings \cite{Colangelo:1996ay,Buchalla:2000sk,Altmannshofer:2009ma}.

The three plots in Fig.~\ref{fig:BsmuvsSphiZprimeA} show rather spectacular differences between these 
three NP scenarios, that can be distinguished from each other provided the 
departures from SM values are sufficiently large. For instance in the case of 
the branching ratio for $B_s\to\mu^+\mu^-$ being smaller than its SM value, $H^0$ 
scenario would be excluded. The fact that there is no overlap between SM prediction and the allowed range in $A^0$ case is related to the requirement of suppressing $\Delta M_s$ below its SM value in order to obtain better agreement with 
experiment. In the $Z^\prime$ case the structure is still different and the 
latter requirement implies a non-vanishing CP-asymmetry $S^s_{\mu\mu}$ which vanishes in the SM. See  \cite{Buras:2013rqa} for corresponding plots.

Yet one should warn the reader that the particular pattern of correlations 
between  $S_{\psi\phi}$ and $\overline{\mathcal{B}}(B_{s}\to\mu^+\mu^-)$ seen 
in Fig.~\ref{fig:BsmuvsSphiZprimeA} 
 depends on  whether the SM value for $\Delta M_s$ is above the data as used in 
\cite{Buras:2012jb} or smaller or equal to it. This has been emphasized and 
illustrated  in \cite{Buras:2012dp} in the context of an explicit 3-3-1 model 
and recently analyzed more generally  in view of the new data on 
$B_{s,d}\to\mu^+\mu^-$ and anomalies in $B_d\to K^*\mu^+\mu^-$ in 
\cite{Buras:2013qja}. Let us then briefly summarize the main findings of 
the latter paper.

\begin{figure}[!tb]
 \centering
\includegraphics[width = 0.45\textwidth]{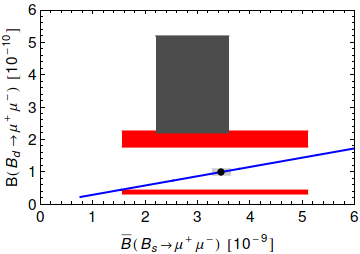}
\includegraphics[width = 0.45\textwidth]{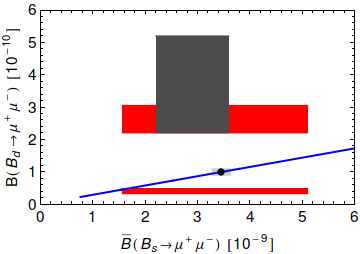}
\caption{ ${\mathcal{B}}(B_{d}\to\mu^+\mu^-)$ versus ${\mathcal{\bar{B}}}(B_{s}\to\mu^+\mu^-)$ in the $Z^\prime$ scenario for $\vub = 0.0034$ (left) and $\vub =
0.0040$ (right) and  $C_{B_d} = 1.04\pm 0.01$, $C_{B_s} = 1.00\pm 0.01$, $\bar{\Delta}_A^{\mu\bar\mu} = 1~\text{TeV}^{-1}$, $0.639\leq
S_{\psi K_s}\leq 0.719$ and $-0.15\leq S_{\psi\phi}\leq 0.15$. SM is represented by the light gray area with black dot and 
 the CMFV prediction by the blue line. Dark gray
 region: Combined exp 1$\sigma$
 range
 $\overline{\mathcal{B}}(B_s\to\mu^+\mu^-) = (2.9\pm0.7)\cdot 10^{-9}$ and $\mathcal{B}(B_d\to\mu^+\mu^-) = (3.6^{+1.6}_{-1.4})\cdot
10^{-10}$.}\label{fig:BdvsBsLHS}~\\[-2mm]\hrule
\end{figure}

\begin{figure}[!tb]
 \centering
\includegraphics[width = 0.45\textwidth]{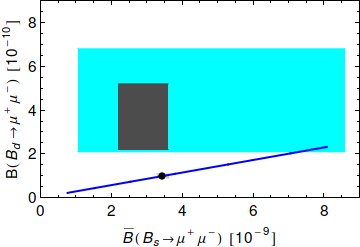}
\includegraphics[width = 0.45\textwidth]{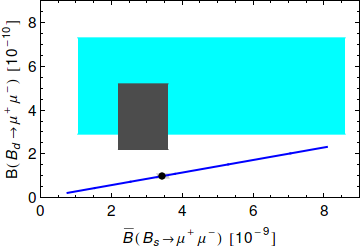}
\caption{ ${\mathcal{B}}(B_{d}\to\mu^+\mu^-)$ versus ${\mathcal{\bar{B}}}(B_{s}\to\mu^+\mu^-)$ in the $Z$-scenario for $\vub = 0.0034$ (left) and $\vub =
0.0040$ (right) and  $C_{B_d} = 0.96\pm 0.01$, $C_{B_s} = 1.00\pm 0.01$, $0.639\leq
S_{\psi K_s}\leq 0.719$ and $-0.15\leq S_{\psi\phi}\leq 0.15$. SM is represented by the light gray area with black dot. Dark gray
 region: Combined exp 1$\sigma$
 range
 $\overline{\mathcal{B}}(B_s\to\mu^+\mu^-) = (2.9\pm0.7)\cdot 10^{-9}$ and $\mathcal{B}(B_d\to\mu^+\mu^-) = (3.6^{+1.6}_{-1.4})\cdot
10^{-10}$.}\label{fig:ZBdvsBsLHS}~\\[-2mm]\hrule
\end{figure}

\section{Tree-Level FCNCs Facing New Data}
In addition to new results on $B_{s,d}\to\mu^+\mu^-$, LHCb collaboration reported new results on angular observables in 
$B_d\to K^*\mu^+\mu^-$ that show significant departures from SM expectations 
\cite{Aaij:2013iag,Aaij:2013qta}. Moreover, new data on the observable $F_L$, 
consistent with LHCb value in \cite{Aaij:2013iag} have been presented by 
CMS \cite{Chatrchyan:2013cda}.
These anomalies in $B_d\to K^*\mu^+\mu^-$ triggered recently 
 two sophisticated analyses \cite{Descotes-Genon:2013wba,Altmannshofer:2013foa} 
with the goal to understand the data and to indicate what type of new physics could be responsible for these departures from the SM. Both analyses point 
toward NP contributions in the 
modified coefficients  $C_{7\gamma}$ and $C_{9}$ with the following shifts with
respect to their SM values: 
\be
C^{\rm NP}_{7\gamma} < 0, \qquad C^{\rm NP}_{9} < 0.
\ee
Other possibilities, in particular involving right-handed currents, have been discussed in \cite{Altmannshofer:2013foa}. We are 
looking forward to the analysis of the authors of \cite{Beaujean:2012uj,Bobeth:2012vn} in order  to see whether some consensus about the size of anomalies in question 
between these three groups has been reached. References to earlier papers on  $B\to K^*\mu^+\mu^-$
 by all these authors can be found in  \cite{Descotes-Genon:2013wba,Altmannshofer:2013foa,Bobeth:2012vn} and \cite{Buras:2013ooa}.

\begin{figure}[!tb]
 \centering
\includegraphics[width = 0.43\textwidth]{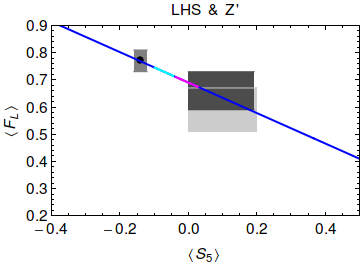}
\includegraphics[width = 0.45\textwidth]{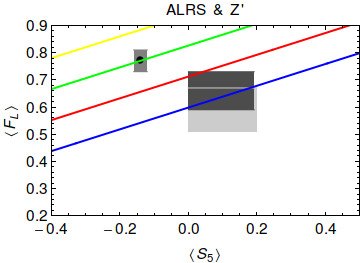}
\caption{Left: $\langle F_L\rangle$ versus $\langle S_5\rangle$ in LHS  where the magenta line corresponds to $C^\text{NP}_9 = -1.6\pm0.3$ and the cyan line to $C^\text{NP}_9 = -0.8\pm0.3$. Right: The same in ALRS 
for different values of $C_9^\text{NP}$: $-2$
(blue), $-1$ (red), $0$ (green) and $1$ (yellow).
The light and dark gray area corresponds to the experimental range for 
$\langle F_L\rangle$ with all data  and only LHCb+CMS data, taken into account,
respectively.  The black point and the
gray box correspond to the SM predictions from \cite{Altmannshofer:2013foa}.}\label{fig:pFLS5LHS}~\\[-2mm]\hrule
\end{figure}

\begin{figure}[!tb]
 \centering
\includegraphics[width = 0.45\textwidth]{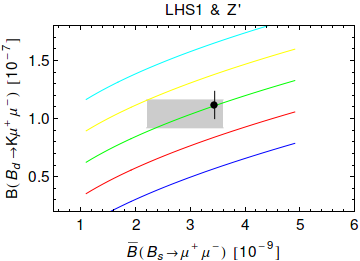}
\caption{$\mathcal{B}(B_d\to K\mu^+\mu^-)$ versus ${\mathcal{\bar{B}}}(B_{s}\to\mu^+\mu^-)$ in LHS for different values of $C_9^\text{NP}$:
$-2$
(blue), $-1$ (red), $0$ (green), $1$ (yellow) and $2$ (cyan) and $-0.8\leq C_{10}^\text{NP}\leq 1.8$. The  gray area corresponds to the
experimental
range. SM is represented by the black point.}\label{fig:pBsmuvsBdKmu}~\\[-2mm]\hrule
\end{figure}

It should be emphasized at this point that these analyses are subject 
to theoretical uncertainties, which have been discussed at length in 
\cite{Khodjamirian:2010vf,Beylich:2011aq,Matias:2012qz,Jager:2012uw,Descotes-Genon:2013wba,Hambrock:2013zya} and it remains to be seen whether the observed anomalies are only 
result of statistical fluctuations and/or underestimated error uncertainties. 
Assuming that this is not the case we have investigated in \cite{Buras:2013qja} 
whether tree-level $Z^\prime$ and $Z$-exchanges could simultaneously 
explain the  $B_d\to K^*\mu^+\mu^-$ anomalies and the most recent data on 
$B_{s,d}\to\mu^+\mu^-$. In this context we have investigated the correlation 
between these decays and $\Delta F=2$ observables. The outcome of this 
rather extensive analysis can be briefly summarized as follows:
\begin{itemize}
\item
The so-called LHS scenario for $Z^\prime$ or $Z$ FCNC couplings (only left-handed quark couplings are flavour violating) provides
a simple model that allows for the violation of the CMFV relation
between the branching ratios for $B_{d,s}\to \mu^+\mu^-$ and $\Delta M_{s,d}$. 
The plots in Figs.~\ref{fig:BdvsBsLHS} and \ref{fig:ZBdvsBsLHS} for $Z^\prime$ 
and $Z$ illustrate this.
\item
However, to achieve this in the case of $Z^\prime$ the experimental value 
of $\Delta M_s$ must be 
very close to its SM value and $\Delta M_d$ is favoured to be by $5\%$ 
{\it larger} than $(\Delta M_d)_{\rm SM}$. $S_{\psi\phi}$ can still deviate 
significantly from its SM value.
\item
In the case of $Z$, both  $\Delta M_s$ and $S_{\psi\phi}$ must be rather close 
to their SM values while $\Delta M_d$ is favoured to be by $5\%$ 
{\it smaller} than $(\Delta M_d)_{\rm SM}$.
\item
As far as  the anomalies in $B\to K^*\mu^+\mu^-$ are concerned $Z^\prime$ 
with only left-handed couplings is capable of softening the anomalies in 
the observables $F_L$ and $S_5$ in a correlated manner as proposed 
 \cite{Descotes-Genon:2013wba,Altmannshofer:2013foa}. However, a better 
description of the present data is obtained by including also right-handed 
contributions with the RH couplings of approximately the same magnitude 
but opposite sign. This is so-called ALRS scenario of \cite{Buras:2012jb}.
We illustrate this in Fig.~\ref{fig:pFLS5LHS}. This is in agreement with the findings in \cite{Altmannshofer:2013foa}. Several analogous correlations can be found in \cite{Buras:2013qja}. We should emphasize that if $Z^\prime$ is the 
only new particle at scales $\ord(\tev)$ than $C^{\rm NP}_{7\gamma}$ can be 
neglected implying nice correlations shown in  Fig.~\ref{fig:pFLS5LHS}.
\item
Strict correlation between  ${\mathcal{\bar{B}}}(B_{s}\to\mu^+\mu^-)$  
and the branching ratio for $B_d\to K\mu^+\mu^-$ at high $q^2$ as a function of $C_9^{\text NP}$ in LHS has been found \footnote{We thank David Straub for pointing it out.}. We show it in  Fig.~\ref{fig:pBsmuvsBdKmu}. The error in the SM prediction for $\mathcal{B}(B_d\to K\mu^+\mu^-)$ is in the ballpark of $10\%$ but the lattice calculations will certainly decrease it with time
\cite{Bouchard:2013mia,Bouchard:2013eph}. This error  should 
 be taken into account in the lines corresponding to NP predictions with 
$C_9^{\rm NP}\not=0$.
Indeed in agreement with  \cite{Altmannshofer:2013foa}  
only $|C_9^{\rm NP}|\le 1.0$ is allowed at $1\sigma$  which is 
insufficient, as seen in Fig.~\ref{fig:pFLS5LHS}, to remove completely 
$B_d\to K^*\mu^+\mu^-$ anomalies in LHS. In ALRS NP contributions to 
$\mathcal{B}(B_d\to K\mu^+\mu^-)$ vanish.
\item
The SM $Z$ boson with FCNC couplings to quarks cannot describe
the anomalies in $B\to K^*\mu^+\mu^-$ due to
its small vector coupling to muons. 
\end{itemize}
 
Finally, let us emphasize that NP effects 
in $\kpn$ and $\klpn$ can be very large in both $Z^\prime$ and $Z$ scenarios 
but are bounded by the upper bound on $K_L\to \mu^+\mu^-$ \cite{Buras:2012jb}.

\section{Towards the new SM in 12 Steps and DNA-Charts}
\begin{figure}[!tb]
\centering
\includegraphics[width=0.65\textwidth]{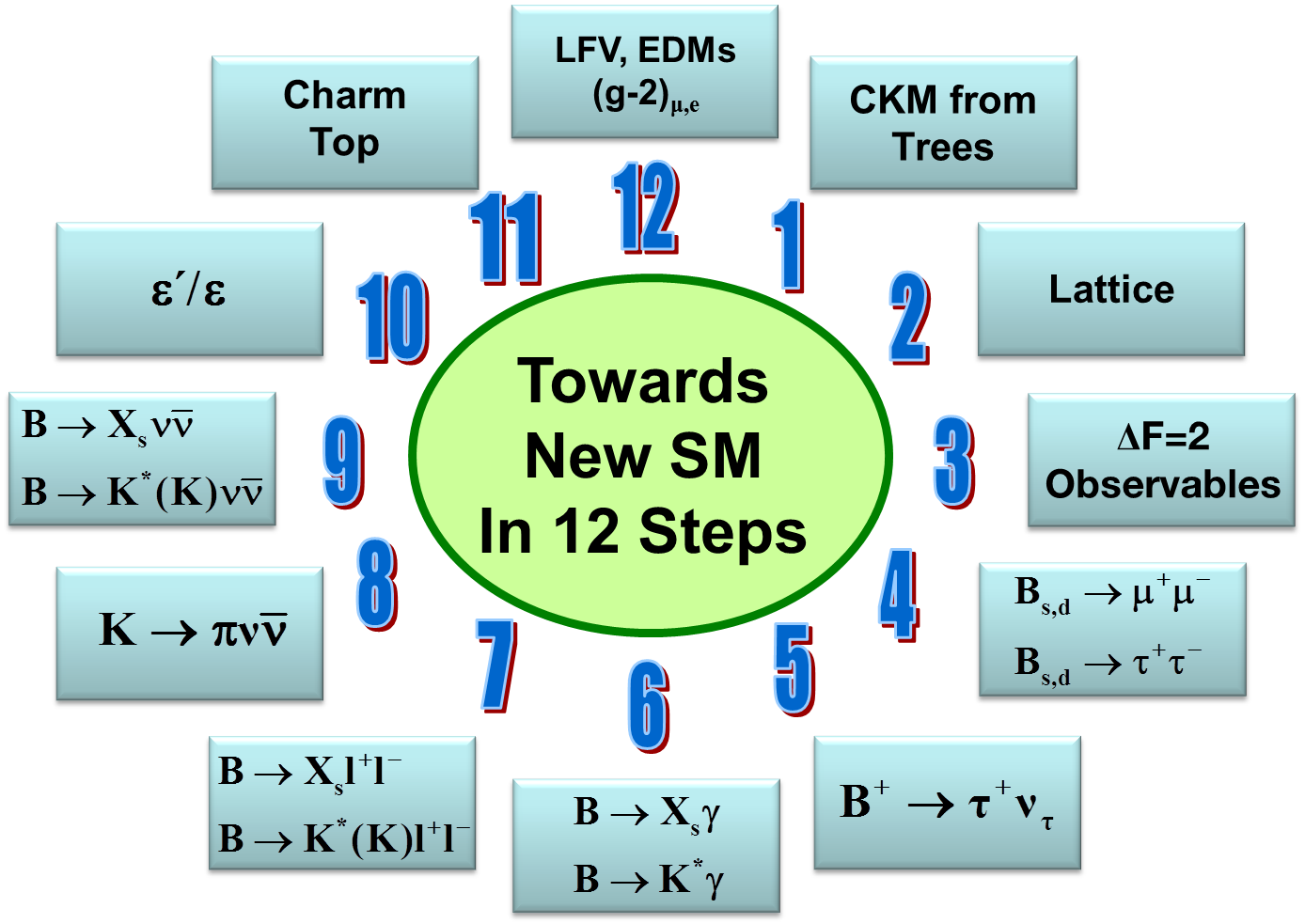}
\caption{\it Towards New Standard Model in 12 Steps.}\label{Fig:1}~\\[-2mm]\hrule
\end{figure}

The identification of NP indirectly will require many measurements. The most 
important are shown in Fig.~\ref{Fig:1} taken from \cite{Buras:2013ooa}, where we have outlined a strategy 
consisting of 12 steps for identifying the correct extension of the SM. 
Very important are the first two steps which should allow to obtain precise 
predictions for the observables considered in the remaining 10 steps within the 
SM. Finding the deviations from SM predictions for these observables in future 
measurements performed in this decade and studying correlations between these 
deviations should allow at least to identify some routes to be followed which
 one day could bring us to the Zeptouniverse. However, as we stressed above,
the pattern of deviations from SM predictions depends crucially on the outcome 
of first two steps. The analyses in \cite{Buras:2013qja,Buras:2013raa} show this 
in explicit terms.

\begin{figure}[!tb]
\centering
\includegraphics[width = 0.65\textwidth]{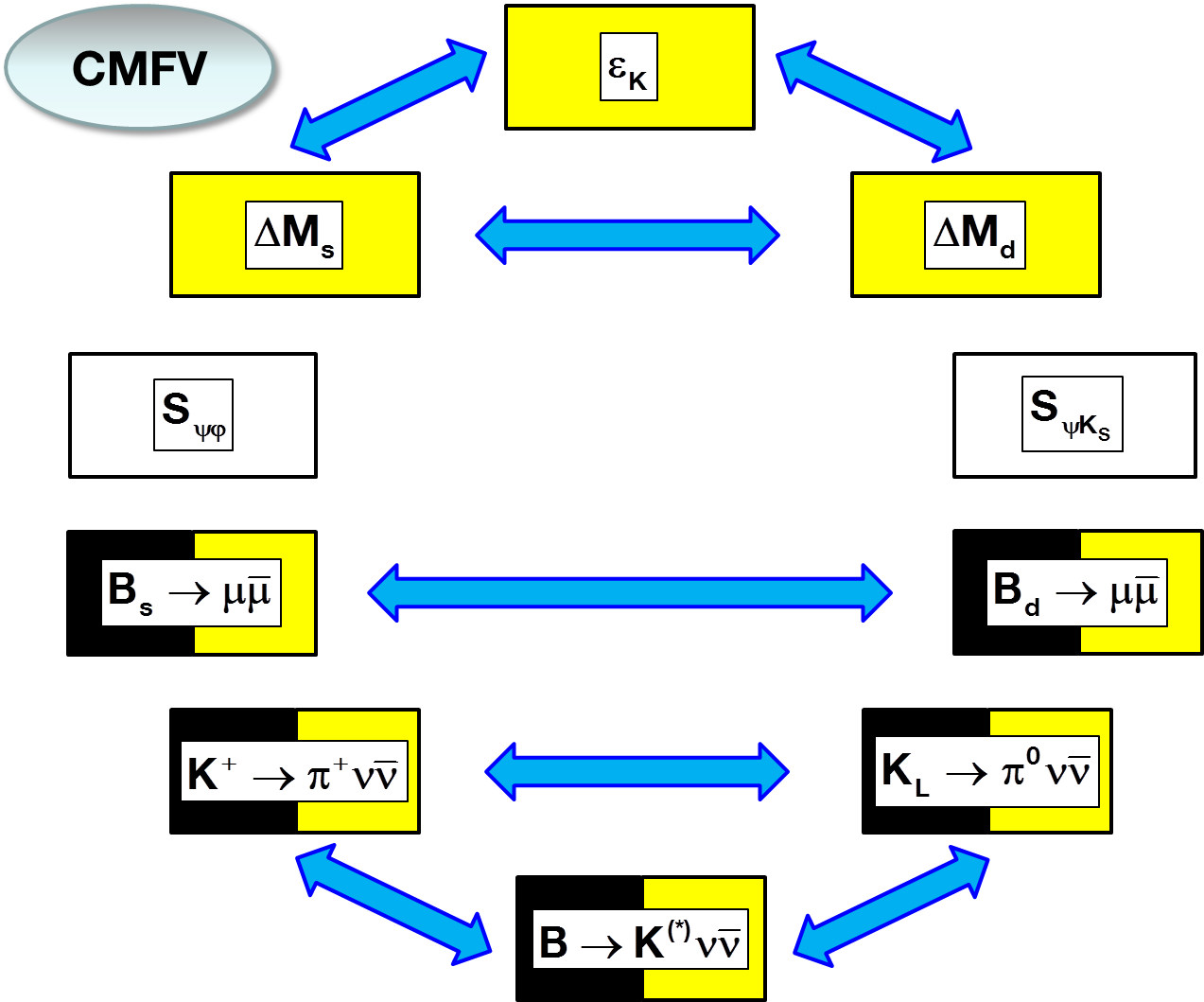}
\caption{\it DNA-chart of CMFV  models. Yellow means   \colorbox{yellow}{enhancement}, black means
\colorbox{black}{\textcolor{white}{\bf suppression}} and white means \protect\framebox{no change}. Blue arrows
\textcolor{blue}{$\Leftrightarrow$}
indicate correlation and green arrows \textcolor{green}{$\Leftrightarrow$} indicate anti-correlation. }
 \label{fig:CMFVchart}~\\[-2mm]\hrule
\end{figure}

\begin{figure}[!tb]
\centering
\includegraphics[width = 0.65\textwidth]{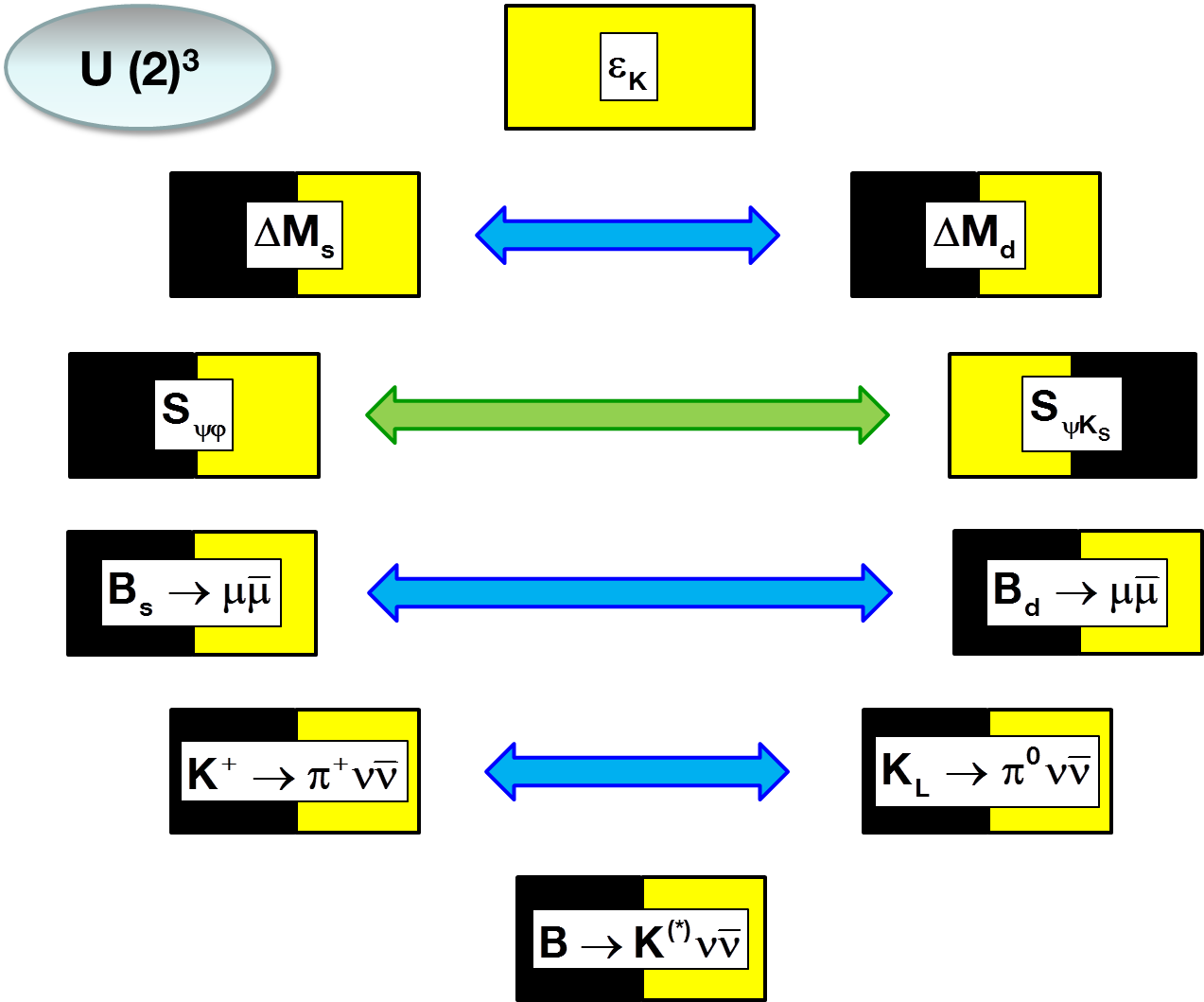}
\caption{\it DNA-chart of $U(2)^3$ models.  Yellow means   \colorbox{yellow}{enhancement}, black means
\colorbox{black}{\textcolor{white}{\bf suppression}} and white means \protect\framebox{no change}. Blue arrows
\textcolor{blue}{$\Leftrightarrow$}
indicate correlation and green arrows \textcolor{green}{$\Leftrightarrow$} indicate anti-correlation. }
 \label{fig:U23chart}~\\[-2mm]\hrule
\end{figure}

\begin{figure}[!tb]
\centering
\includegraphics[width = 0.49\textwidth]{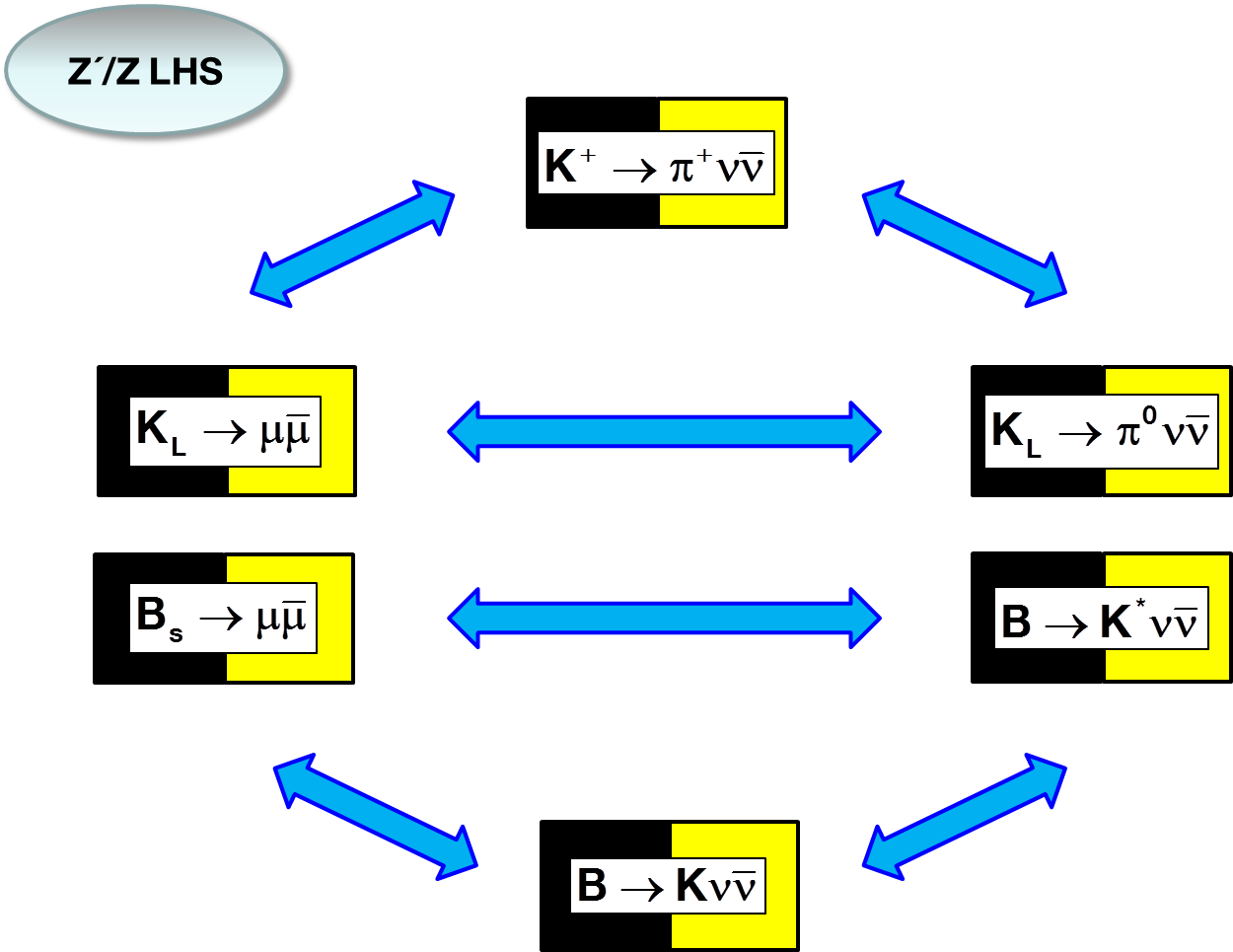}
\includegraphics[width = 0.49\textwidth]{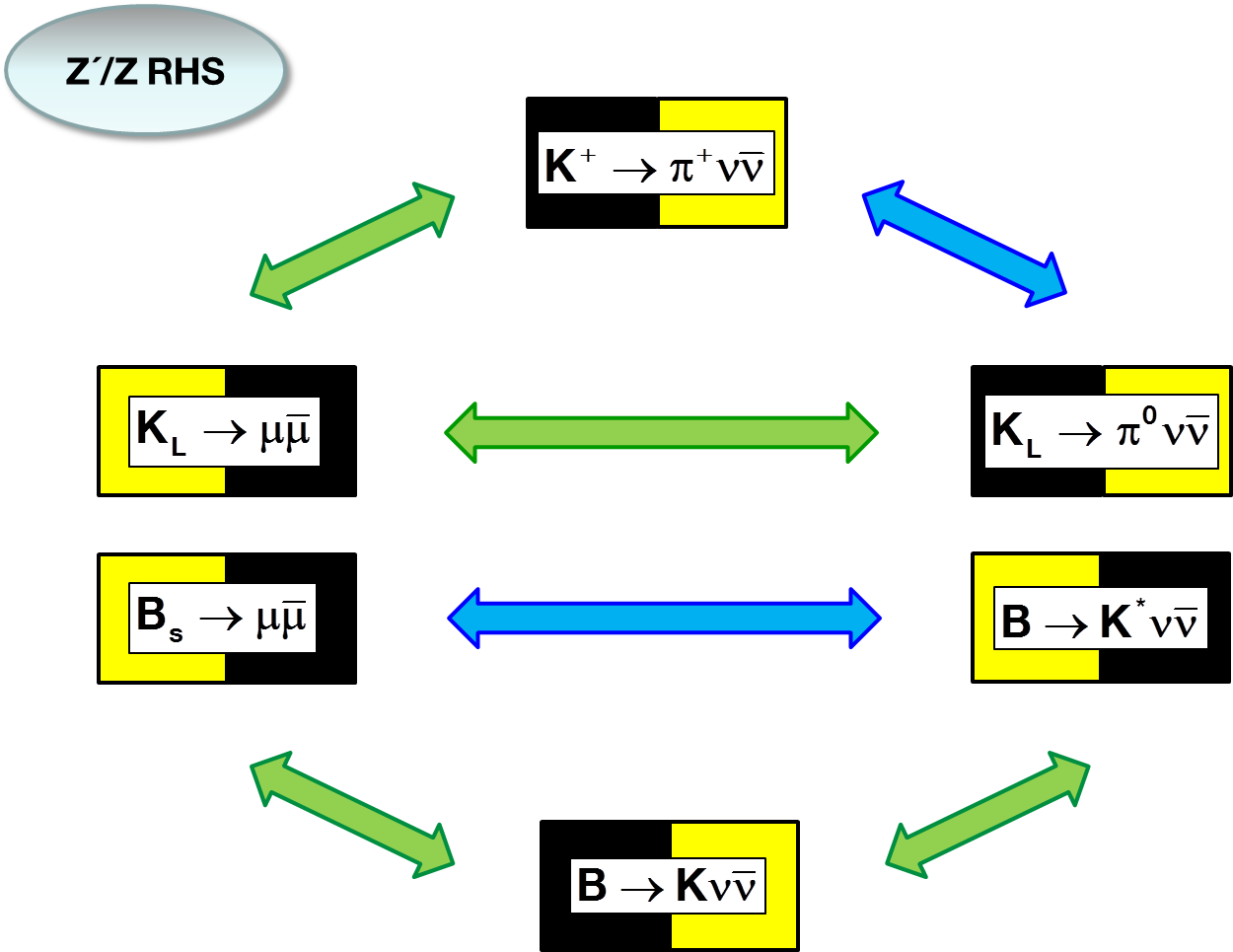}

\caption{\it DNA-charts of $Z^\prime$ models with LH and RH currents.  Yellow means   \colorbox{yellow}{enhancement}, black means
\colorbox{black}{\textcolor{white}{\bf suppression}} and white means \protect\framebox{no change}. Blue arrows
\textcolor{blue}{$\Leftrightarrow$}
indicate correlation and green arrows \textcolor{green}{$\Leftrightarrow$} indicate anti-correlation. }
 \label{fig:ZPrimechart}~\\[-2mm]\hrule
\end{figure}

As emphasized in \cite{Buras:2013ooa} already the pattern of signs of departures  from SM expectations in various observables and  the correlations or anti-correlations between these
departures could exclude or 
support certain NP scenarios. In order to depict various possibilities in 
a transparent manner a DNA-Chart has been proposed to be applied separately 
to each NP scenario. 
In Fig.~\ref{fig:CMFVchart} we show the DNA-chart of CMFV and the corresponding chart for $U(2)^3$ models
is shown in Fig.~\ref{fig:U23chart}. 
The DNA-charts representing models with left-handed and right-handed flavour violating
couplings of  $Z$ and $Z^\prime$  can be found in Fig.~\ref{fig:ZPrimechart}.

The interested reader may check that these charts summarize compactly the 
correlations that are discussed in detail at various places in \cite{Buras:2013ooa}. In particular we observe the following features:
\begin{itemize}
\item
When going from  the DNA-chart of CMFV in  Fig.~\ref{fig:CMFVchart} to the 
one for the $U(2)^3$ models in  Fig.~\ref{fig:U23chart}, the correlations 
between $K$ and $B_{s,d}$ systems are broken as the symmetry is reduced from 
$U(3)^3$ down to $U(2)^3$. The anti-correlation between $S_{\psi\phi}$ and 
$S_{\psi K_S}$ is just the one shown in Fig.~\ref{fig:SvsS}.
\item
As the decays $\kpn$, $\klpn$ and $B\to K\nu\bar\nu$  are only sensitive
to the vector quark currents, they do not change when the couplings are changed from  left-handed to right-handed ones. On the other hand the remaining 
three decays in   Fig.~\ref{fig:ZPrimechart} are sensitive to axial-vector 
couplings implying interchange of enhancements and suppressions when going from 
$L$ to $R$ and also change of correlations to anti-correlations between the 
latter three and the former three decays. Note that the correlation between 
$B_s\to\mu^+\mu^-$  and $B\to K^*\mu^+\mu^-$ does not change as both decays are  sensitive only to axial-vector coupling. 
\item
However, it should be remarked that in order to obtain the correlations or 
anti-correlations in LHS and RHS scenarios it was assumed in the DNA charts 
presented here that the signs 
of the left-handed couplings to neutrinos and the axial-vector couplings 
to muons are the same which does not have to be the case. If they are 
opposite the correlations between the decays with neutrinos and muons in 
the final state change to anti-correlations and vice versa. 
\item
On the other hand due to $SU(2)_L$ symmetry the left-handed $Z^\prime$
 couplings to muons and neutrinos are equal and this implies the relation
\be\label{SU2}
\Delta_{L}^{\nu\bar\nu}(Z')=\frac{\Delta_V^{\mu\bar\mu}(Z')-\Delta_A^{\mu\bar\mu}(Z')}{2}. 
\ee
Therefore, once two of these couplings are determined the third follows uniquely without the freedom mentioned in the previous item.
\item
In the context of the DNA-charts in  Fig.~\ref{fig:ZPrimechart}, the correlations involving $\klpn$ apply only if NP contributions carry some CP-phases. If this is not the case the branching ratio for $\klpn$ will remain unchanged relativ to the SM one.
\end{itemize}

If in the case of tree-level $Z^\prime$ and $Z$ exchanges 
both LH and RH quark couplings are present which in addition are equal to each 
other (LRS scenario) or differ by sign (ALRS scenario) then one finds 
\cite{Buras:2012jb}
\begin{itemize}
\item
In LRS NP contributions to $B_{s,d}\to\mu^+\mu^-$ vanish but not to $\klpn$, 
$\kpn$ and $B_d\to K\mu^+\mu^-$.
\item
In ALRS NP contributions to $B_{s,d}\to\mu^+\mu^-$ are non-vanishing and 
this also applies to $B_d\to K^*\mu^+\mu^-$ as seen in the right panel 
of Fig.~\ref{fig:pFLS5LHS}. On the other hand 
they vanish in the case of  $\klpn$, $\kpn$ and  $B_d\to K\mu^+\mu^-$.
\end{itemize}

In summary there are exciting times ahead of us and following the 12 Steps in 
Fig.~\ref{Fig:1} and studying correlations between various observables we may one day reach the Zeptouniverse.

{\bf Acknowledgements}\\
I would like to thank first of all Jennifer Girrbach for numerous studies 
of various aspects of NP models which resulted in the review in 
\cite{Buras:2013ooa}, in the 12 steps in Fig.~\ref{Fig:1} and  DNA charts  just discussed.
I also thank all other collaborators for exciting time we spent together 
exploring the short distance scales with the help of flavour violating 
processes. Finally I would like to thank the organizers of the flavour session for being generous with the time given to this talk. 
This research was dominantly financed and done in the context of the ERC Advanced Grant project ``FLAVOUR'' (267104) and carries the number ERC-49. It was also partially supported by the 
DFG cluster of excellence ``Origin and Structure of the Universe''.



\end{document}